\documentstyle[multicol,aps,prc,epsf]{revtex}

\draft
\begin{document}

\title{Comment on \lq\lq Tensor force in doubly odd deformed nuclei\rq\rq}
\author{D.\ Nosek$^{1}$ %\thanks{nosek@hp02.troja.mff.cuni.cz} 
        and J.\ Noskov\' a$^{2}$ %\thanks{noskova@mat.fsv.cvut.cz}
        }
\address{$^{1}$Department of Nuclear Physics, Charles University,
         V Hole\v sovi\v ck\'ach 2, 180 00 Prague, Czech Republic}
\address{$^{2}$Department of Mathematics, Faculty of Civil Engineering,
         Czech Technical University in Prague, \\
         Th\' akurova 7, 166 29 Prague, Czech Republic}
\date{\today}

\maketitle

\begin{abstract}
An article by Covello, Gargano and Itaco 
$\lbrack$Phys.\ Rev.\ C {\bf 56}, 3092 (1997)$\rbrack$ 
tries to find evidence for the important role of the residual 
tensor force between the valence proton and neutron in doubly 
odd deformed nuclei.
It is shown that observable effects discussed by these authors 
do not fully justify their rather strong conclusions. 
\end{abstract}

\pacs{PACS numbers: 21.60.Ev, 27.70.+q, 27.90.+b} 

\begin{multicols}{2}

Recently, the role of the effective proton--neutron (p--n) \linebreak 
interaction with particular attention focused on the tensor force operating
between the unpaired proton and neutron in odd--odd deformed nuclei 
was studied by Covello~et~al.~\cite{Cov1}.
Their results obtained for $^{176}$Lu were interpreted as evidence
for the originally proposed central+tensor force~\cite{Boi1}
which \lq\lq may be profitably used for a systematic study of doubly 
odd deformed nuclei in the rare--earth region\rq\rq~\cite{Cov1}.
In this Comment we critically analyze rather strong conclusions 
articulated by Covello~et~al.~\cite{Cov1} and also by Itaco~et~al.~\cite{Cov2}.
Our method and results~\cite{Nos1}, that was also questioned in 
Ref.~\cite{Cov1}, will be defended elsewhere.

In their study, Covello~et~al.~\cite{Cov1} carried out calculations 
of the spectrum of the first K$^{\pi}$=0$^{-}$ band built upon the   
\{$\frac{7}{2}$+[404]~$\frac{7}{2}$--[514]\} intrinsic
configuration in $^{176}$Lu and found out that very good agreement 
with experiment is achieved only if the central+tensor forces with 
the Gaussian radial shape are taken into account when the Newby (N) 
shift in this band is estimated. 
They stated that its empirical value is exactly reproduced when 
the p--n parameters as recommended by Boisson~et~al.~\cite{Boi1} are used. 
Then, not surprisingly, also the whole rotational band is very well
reproduced since there is no evidence of mixing with other bands.
But, it is interesting to note that this N shift was not
satisfactorily calculated in Ref.~\cite{Boi1} even if the tensor 
terms of the Gaussian force were set active, see Table~\ref{Tab1}. 
On the other hand, our finite--range values of this N shift the
estimates of which are based on different empirical sets 
of the p--n parameters~\cite{Nos1} agree well with experiment 
as shown in Table~\ref{Tab1}. 
Consequently, also our empirical sets of the p--n parameters with not
well determined tensor strengths yield a consistent explanation 
of the spectrum of the lowest K$^{\pi}$=0$^{-}$ band.
Unfortunately, the authors of Ref.~\cite{Cov1} did not provide
any comparison of their theoretical B$_{N}$ values in this band 
with our results~\cite{Nos1}.

In our opinion, there is more serious difficulty concerning the 
predictive power of the experimental spectrum of the first K$^{\pi}$=0$^{-}$ 
rotational band in $^{176}$Lu. 
Since the relevant N shift is of the central type (NC
shift~\cite{Boi1}), it is not expected to possess significant
tensor contributions~\cite{Boi1}.
In our analysis the results of which are given in Table~\ref{Tab1}, 
the central terms of the Gaussian force are sufficient to predict 
its reasonable theoretical value even though small tensor
contributions with the right signs are calculated. 
Thus, this particular example have a little to do with 
the importance of the tensor--force effects. 

It is worth noting that the N shift measured in the K$^{\pi}$=0$^{+}$ 
\{$\frac{7}{2}$+[404]~$\frac{7}{2}$+[633]\} band
in $^{174}$Lu and also in $^{170,172}$Lu and $^{176}$Ta~\cite{Nos1}
which is of the tensor type (NT shift~\cite{Boi1}) gives a better 
picture of the tensor--force effects.
Although the strengths of the tensor forces are not well determined 
in our experimental set of the N shifts~\cite{Nos1}, their effect is 
well visible in Table~\ref{Tab1} where different values of this N
shift are collected. 
(Notice that our theoretical B$_{N}$ values are very different from 
those calculated in Ref.~\cite{Cov2}.)
In our calculations, similarly as in the previous example, the 
finite--range tensor forces operate in the right direction.
Their contributions are, however, more significant since the central 
forces alone do not provide an acceptable B$_{N}$ value.
Nonetheless, the tensor--force effects are rather small in order that one
can deduce a definite conclusion.

The second point we want to discuss is the role of 
irregularities which are known
to be present in rotational bands in $^{176}$Lu.
In Ref.~\cite{Cov1} and a subsequent preprint~\cite{Cov2}, the
authors analyzed an odd--even staggering observed experimentally in 
the two lowest K$^{\pi}$=1$^{+}$ rotational bands built
upon the \{$\frac{9}{2}$--[514]~$\frac{7}{2}$--[514]\} and 
\{$\frac{7}{2}$+[404]~$\frac{9}{2}$+[624]\} 
configurations, respectively.
They suggested that this rather large staggering may be caused by direct
Coriolis coupling with Newby--shifted K$^{\pi}$=0$^{+}$ bands 
assigned as \{$\frac{7}{2}$--[523]~$\frac{7}{2}$--[514]\} and
\{$\frac{7}{2}$+[404]~$\frac{7}{2}$+[633]\}, respectively.
But these effects are small; typically 5--15~\% admixtures were 
reported in Refs.~\cite{Cov1,Cov2}.
In the former case, the main trends were reproduced with the 
central+tensor Gaussian force~\cite{Cov1}.
The latter effect, that is equally well developed in the experimental
spectrum, failed to be described satisfactorily.
The best picture, even if it is hardly acceptable, 
was obtained with the same type of the residual p--n force~\cite{Cov2}.
All these results were then interpreted as \lq\lq clear evidence of the
importance of the tensor--force effects\rq\rq~\cite{Cov1},
see also Ref.~\cite{Cov2}.

\end{multicols}
\begin{center}
%%%%%%%%%%%%%%%%%%%%%%%%%%%%%%%%%%%%%%%%%%%%%%%%%%%%%%%%%%%%%%%%%%%%%%%%%%%
\begin{table}[t!]
\caption{Newby shifts, B$_{N}$, in the three discussed
  K$^{\pi}$=0$^{\pm}$ bands in $^{174,176}$Lu. 
  Their types are given in the third column.
  The experimental and theoretical central+tensor (CT) values obtained 
  by Boisson~et~al.~\protect\cite{Boi1} are written in the forth and
  fifth columns, respectively.
  The theoretical values obtained by Covello et al.~\protect\cite{Cov1}
  and Itaco et al.~\protect\cite{Cov2} with 
  the $\delta$ force ($\delta$), the central (C) and central+tensor
  (CT) Gaussian force are listed in the sixth, seventh and eighth
  columns, respectively. 
  Our empirical and theoretical values of these N shifts, the latter  
  obtained in different fits~\protect\cite{Nos1} with the $\delta$
  potential ($\delta$), the central (G) and central+tensor (G$_{T}$) 
  Gaussian force, and with the intrinsic spin polarization effects 
  (fits $\delta_{P}$, G$_{P}$ and G$_{TP}$), are given in the following 
  seven columns. }

\begin{tabular}{c c c c c c c c c c c c c c c}
 & & & \multicolumn{12}{c}{B$_{N}$ [keV]} \\[1mm]
\cline{4-15} \\[-3mm]
 & & & \multicolumn{2}{c}{Ref.~\cite{Boi1}} &
       \multicolumn{3}{c}{Refs.~\cite{Cov1,Cov2}} &
       \multicolumn{7}{c}{Present values} \\[1mm]
\cline{4-15} \\[-3mm]
Nucleus & 
\{$\Omega$$\pi$[Nn$_{z}$$\Lambda$]$_{p}$~
$\Omega$$\pi$[Nn$_{z}$$\Lambda$]$_{n}$\} & Type &
 Exp & CT & $\delta$ & C & CT &
Exp & $\delta$ & $\delta_{P}$ & G & G$_{T}$ & G$_{P}$ & G$_{TP}$ \\[1mm]
\hline\\[-3mm]
$^{176}$Lu & \{$\frac{7}{2}$+[404]~$\frac{7}{2}$--[514]\} & NC & 
 \ 69 & \ 35 & -- & -- & \ 70 & 
\ \ 69.2(0.6) & \ 12.1 & \ 32.9 & \ 78.2 & \ 64.8 & \ 73.5 & \ 64.6 \\[1mm]
$^{174}$Lu & \{$\frac{7}{2}$+[404]~$\frac{7}{2}$+[633]\} & NT &  
 -35 & -29 & 6 & 1 & -26 & 
\ -40.0(3.2) & -27.1 & -43.2 & -28.2 & -35.7 & -32.5 & -36.5 \\[1mm]
$^{176}$Lu & \{$\frac{7}{2}$\,--[523]~$\frac{7}{2}$\,--[514]\} & NT & 
 -- & -- & -- & -- & -- & 
-155.5(1.2) & -28.7 & \ \ 4.0 & -22.6 & -40.7 & -18.7 & -35.4 
\end{tabular}
\label{Tab1}
\end{table}  
%%%%%%%%%%%%%%%%%%%%%%%%%%%%%%%%%%%%%%%%%%%%%%%%%%%%%%%%%%%%%%%%%%%%%%%%%%%
\vspace{-2mm}
%%%%%%%%%%%%%%%%%%%%%%%%%%%%%%%%%%%%%%%%%%%%%%%%%%%%%%%%%%%%%%%%%%%%%%%%%%%
\begin{table}[t!]
\caption{Gallagher-Moszkowski splitting energies, $\Delta$E$_{GM}$, 
  for the two discussed intrinsic configurations with
  K$^{\pi}$=1$^{+}$ and 8$^{+}$ in $^{176}$Lu. 
  The experimental and theoretical GM values given by other 
  authors~\protect\cite{Cov1,Boi1} as well as the present 
  values based on the revised interpretation~\protect\cite{Kla1} 
  are summarized.
  Our present experimental values of the GM splitting energies written 
  in the ninth column are corrected for the $\Delta$K=0 interaction, 
  for more detail see text. }

\begin{tabular}{c c c c c c c c c c c c c}
 & & \multicolumn{11}{c}{$\Delta$E$_{GM}$ [keV]} \\[1mm]
\cline{3-13} \\[-3mm]
 & &
\multicolumn{3}{c}{Ref.~\cite{Boi1}} &
\multicolumn{2}{c}{Ref.~\cite{Cov1}} & 
\multicolumn{6}{c}{Present values} \\[1mm]
\cline{3-13} \\[-3mm]
Nucleus & 
\{$\Omega$$\pi$[Nn$_{z}$$\Lambda$]$_{p}$~
$\Omega$$\pi$[Nn$_{z}$$\Lambda$]$_{n}$\} & 
 Exp & C & CPTL & Exp & CT & Exp & Exp $\Delta$K=0 & 
G & G$_{T}$ & G$_{P}$ & G$_{TP}$ \\[1mm]
\hline\\[-3mm]
$^{176}$Lu & \{$\frac{9}{2}$\,--[514]~$\frac{7}{2}$\,--[514]\} &
 -- & -239 & -141 & -219 & -154 & -220.3(5.1) & -223.9(5.2) & 
-184.3 & -200.8 & -168.3 & -178.4 \\[1mm] 
$^{176}$Lu & \{$\frac{7}{2}$+[404]~$\frac{9}{2}$+[624]\} &
 -- & -130 & -107 & \ -12 & \ -90 & \ -15.4(3.2) & \ -51.8(3.2) & 
-159.9 & -132.4 & -150.1 & -117.7 
\end{tabular}
\label{Tab2}
\end{table}
%%%%%%%%%%%%%%%%%%%%%%%%%%%%%%%%%%%%%%%%%%%%%%%%%%%%%%%%%%%%%%%%%%%%%%%%%%%
\end{center}
\begin{multicols}{2}

In our previous study~\cite{Nos1}, we have assumed an old
interpretation~\cite{Bal1,Dew1} for the two lowest K$^{\pi}$=1$^{+}$ 
rotational bands with the band heads at 194 keV and 338 keV, respectively.
The revised interpretation of these bands given by
Klay~et~al.~\cite{Kla1} yields theoretical values of 
the Gallagher--Moszkowski (GM) splitting energies calculated 
with our sets of the p--n parameters which are in even better
agreement with experiment than our previous results~\cite{Nos1}, 
see Table~\ref{Tab2}.

We have carried out preliminary calculations of the spectrum
of $^{176}$Lu assuming 42 low--lying rotational bands of positive
parities and including the Coriolis interaction, intrinsic rotational
contributions, recoil terms, diagonal terms of the residual p--n
interaction, and also non--diagonal p--n mixing ($\Delta$K=0
interaction); the latter was not considered in Refs.~\cite{Cov1,Cov2}.
We have used the same mean--field parameters of the Nilsson potential 
as in Ref.~\cite{Nos1}. 
The G$_{T}$ sets of the p--n parameters have been adopted from the same
study.

The most striking feature of our calculations is that there is 
strong $\Delta$K=0 mixing between both K$^{\pi}$=1$^{+}$ rotational 
bands as well as between their K$^{\pi}$=8$^{+}$ GM partners with 
non--diagonal matrix elements $\mid < V_{pn} > \mid \approx 50-60$~keV.  
When included, this interaction, for example, affects significantly 
empirical values of the GM splitting energies.
This is demonstrated in Table~\ref{Tab2} where corrected empirical 
values, that are obtained assuming the G$_{T}$(GM) set~\cite{Nos1} of 
the p--n parameters for the $\Delta$K=0 interaction, are written 
in the ninth column.
This finding suggests that the odd--even staggering can be transferred
from one band to the other; such a picture was not confirmed  
in Refs.~\cite{Cov1,Cov2}.
Further, since the
K$^{\pi}$=0$^{+}$~\{$\frac{7}{2}$--[523]~$\frac{7}{2}$--[514]\} 
band is expected to lie very high in energy (its band head was 
tentatively placed at 1057~keV~\cite{Dew1}), 
its influence on the low--lying K$^{\pi}$=1$^{+}$ 
band is found to be smaller in our calculations than in those
performed in Ref.~\cite{Cov1}. 
In particular, the odd--even staggering that is discussed in 
Ref.~\cite{Cov1}, is equally well explained in our calculations 
only if very strong Coriolis mixing with the relevant
K$^{\pi}$=0$^{+}$ band exists. 
It holds if our theoretical G$_{T}$ value of the corresponding 
N shift is taken from Table~\ref{Tab1}.
Let us note that also in this case the tensor--force effects are not 
negligible. 
However, due to tentative assignment of this K$^{\pi}$=0$^{+}$ 
band~\cite{Dew1}, an extremely large absolute empirical B$_{N}$ value 
has to be regarded as very uncertain~\cite{Nos1} and can hardly be
compared with our predictions.
Unfortunately, the authors of Refs.~\cite{Cov1,Cov2} did not provide 
any theoretical value of this quantity.
On the other hand, the odd--even staggering in the 
K$^{\pi}$=1$^{+}$~\{$\frac{7}{2}$+[404]~$\frac{9}{2}$+[624]\} 
band, that is badly described in Ref.~\cite{Cov2}, is very well
reproduced in our calculations based on our well determined value 
of the N shift in the 
K$^{\pi}$=0$^{+}$~\{$\frac{7}{2}$+[404]~$\frac{7}{2}$+[633]\} 
band in $^{174}$Lu, see Table~\ref{Tab1}.
Moreover, we have found that, due to the $\Delta$K=0 interaction, 
the latter staggering is partially transformed into the 
K$^{\pi}$=1$^{+}$ band lying lower in energy.
Nevertheless, a better analysis of this effect is required.

In conclusion, we would like to stress that our previous
statement~\cite{Nos1} that the p--n parameters of the tensor 
forces are not well determined in our set of presently known N shifts
does not imply that their role should be negligible.
Here, we are forced to infer that, although probably right in principle, 
the conclusions concerning the importance of the tensor--force effects 
drawn in Refs.~\cite{Cov1,Cov2} are not sufficiently supported.
The reason is that a particular example (the lowest K$^{\pi}$=0$^{-}$ 
band in $^{176}$Lu) cannot indicate general features which are known 
to be extremely subtle.
Nearly the same picture is obtained when the space--exchange and
spin--spin space--exchange forces~\cite{Nos1} for the description 
of the N shifts are assumed to be the most important.
In such a way, the spectrum of the lowest K$^{\pi}$=0$^{-}$ band does
not provide any argument against our previous conclusions~\cite{Nos1}.

It should be finally pointed out that there remains a place for 
a different explanation of the observed odd--even staggering in 
the low--lying K$^{\pi}$=1$^{+}$ bands discussed in 
Refs.~\cite{Cov1,Cov2}. 
The crucial point for its understanding lies in a proper estimate 
of non--diagonal mixing which is caused by the Coriolis coupling, 
as correctly suggested in Refs.~\cite{Cov1,Cov2}, 
but also by the $\Delta$K=0 interaction.
Thus, we conclude that the odd-even staggering and its theoretical
description including the tensor terms does not directly imply 
that \lq\lq only this force is able to reproduce a sizable N
shift\rq\rq~\cite{Cov1,Cov2} for both K$^{\pi}$=0$^{+}$ bands under 
consideration.

\medskip
%\acknowledgements
This work was supported by the Grant Agency in the Czech Republic
under Contract No.~GA\v CR~202/99/1718.

%%%%%%%%%%%%%%%%%%%%%%%%%%%%%%%%%%%%%%%%%%%%%%%%%%%%%%%%%%%%%%%%%%%%%%%%%%%

%%%%%%%%%%%%%%%%%%%%%%%%%%%%%%%%%%%%%%%%%%%%%%%%%%%%%%%%%%%%%%%%%%%%%%%%%%%
\end{multicols}

\end{document}